\documentclass[a4paper,11pt]{article}
\usepackage{aaskaiid}
\usepackage{subfigure}
\usepackage{orcidlink}
\NewCommandCopy\oldtextbf\textbf

\title{Impact of Anomalous Microwave Emission (AME) on Radio Spectral Energy Distributions: SKA Observations of Galaxies Near and Far}
\ShortTitle{AME impact on radio SED}

\author[1,2]{Ilsang Yoon\orcidlink{0000-0001-9163-0064}}
\ShortName{Yoon et al.} % shortened name list for header 
\author[1,2]{Eric Murphy\orcidlink{0000-0001-7089-7325}}
\author[3]{Caroline Bot\orcidlink{0000-0001-6118-2985}}
\author[3]{Lucie Correia\orcidlink{0000-0003-1225-5820}}

\affiliation[1]{National Radio Astronomy Observatory, 520 Edgemont Road, Charlottesville, VA 22903, USA}
\emailAdd{iyoon@nrao.edu}
\affiliation[2]{Department of Astronomy, University of Virginia, P.O. Box 3818, Charlottesville, VA 22903, USA}
\affiliation[3]{Observatoire astronomique de Strasbourg, Université de Strasbourg, CNRS, UMR 7550, Strasbourg, F-67000, France}

\abstract{
Radio continuum emission powered by the thermal bremsstrahlung process is a clean, dust-free tracer of star formation in galaxies. However, the existence of anomalous microwave emission (AME) that is also prominent in a similar frequency range may challenge the use of thermal radio continuum emission to measure the star formation rates of galaxies. So while the nature of AME and the ISM conditions that lead to strong observable AME are still not well understood, the impact of AME on the radio emission from galaxies needs to be investigated for the SKA, which will be sensitive to large numbers of faint, high-redshift galaxies. In this chapter, we compute the observable flux density of free-free emission and AME and investigate the impact of AME on the galaxy radio spectral energy distribution for given observing frequencies and redshifts. Our conclusion is that (1) significance of AME is determined by the size of the AME region relative to the observing beam and ISM hydrogen column density, (2) thermal free-free emission dominates the radio continuum emission for distant galaxies at $\approx 10$ GHz frequency with negligible contribution from AME, (3) high-angular resolution observation of nearby galaxies resolving individual star forming region may need multi-frequency observations to avoid a potential bias from AME in the measure of star formation rate from single frequency observation.   
}

\begin{document}
\newcommand{\actaa}{Acta Astron.} % Acta Astronomica
\newcommand{\araa}{Annu. Rev. Astron. Astrophys.} % Annual Review of Astron and Astrophys
\newcommand{\aar}{Astron. Astrophys. Rev.} % Astrononmy and Astrophysics Review
\newcommand{\ab}{Astrobiol.} % Astrobiology
\newcommand{\aj}{Astron. J.} % Astronomical Journal
\newcommand{\apj}{Astrophys. J.} % Astrophysical Journal
\newcommand{\apjl}{Astrophys. J. Lett.} % Astrophysical Journal, Letters
\newcommand{\apjs}{Astrophys. J. Suppl. Ser.} % Astrophysical Journal, Supplement
\newcommand{\ao}{Appl. Opt.} % Applied Optics
\newcommand{\apss}{Astrophys. Space Sci.} % Astrophysics and Space Science
\newcommand{\aap}{Astron. Astrophys.} % Astronomy and Astrophysics
\newcommand{\aapr}{Astron. Astrophys. Rev.} % Astronomy and Astrophysics Reviews
\newcommand{\aaps}{Astron. Astrophys. Suppl.} % Astronomy and Astrophysics, Supplement
\newcommand{\baas}{Bull. Am. Astron. Soc.} % Bulletin of the AAS
\newcommand{\caa}{Chinese Astron. Astrophys.} % Chinese Astronomy and Astrophysics
\newcommand{\cjaa}{Chinese J. Astron. Astrophys.} % Chinese Journal of Astronomy and Astrophysics
\newcommand{\cqg}{Class. Quantum Gravity} % Classical and Quantum Gravity
\newcommand{\gal}{Galaxies} % Galaxies
\newcommand{\gca}{Geochim. Cosmochim. Acta} % Geochimica Cosmochimica Acta
\newcommand{\icarus}{Icarus} % Icarus
\newcommand{\jcap}{J. Cosmol. Astropart. Phys.} % Journal of Cosmology and Astroparticle Physics
\newcommand{\jgr}{J. Geophys. Res.} % Journal of Geophysics Research
\newcommand{\jgrp}{J. Geophys. Res.: Planets} % Journal of Geophysics Research: Planets
\newcommand{\jqsrt}{J. Quant. Spectrosc. Radiat. Transf.} % Journal of Quantitiative Spectroscopy and Radiative Transfer
\newcommand{\memsai}{Mem. Soc. Astron. Italiana} % Mem. Societa Astronomica Italiana
\newcommand{\mnras}{Mon. Not. R. Astron. Soc.} % Monthly Notices of the RAS
\newcommand{\nat}{Nature} % Nature
\newcommand{\nastro}{Nat. Astron.} % Nature Astronomy
\newcommand{\ncomms}{Nat. Commun.} % Nature Communications
\newcommand{\nphys}{Nat. Phys.} % Nature Physics
\newcommand{\na}{New Astron.} % New Astronomy
\newcommand{\nar}{New Astron. Rev.} % New Astronomy Review
\newcommand{\physrep}{Phys. Rep.} % Physics Reports
\newcommand{\pra}{Phys. Rev. A} % Physical Review A: General Physics
\newcommand{\prb}{Phys. Rev. B} % Physical Review B: Solid State
\newcommand{\prc}{Phys. Rev. C} % Physical Review C
\newcommand{\prd}{Phys. Rev. D} % Physical Review D
\newcommand{\pre}{Phys. Rev. E} % Physical Review E
\newcommand{\prl}{Phys. Rev. Lett.} % Physical Review Letters
\newcommand{\psj}{Planet. Sci. J.} % Planetary Science Journal
\newcommand{\planss}{Planet. Space Sci.} % Planetary Space Science
\newcommand{\pnas}{Proc. Natl Acad. Sci. USA} % Proceedings of the US National Academy of Sciences
\newcommand{\procspie}{Proc. SPIE} % Proceedings of the SPIE
\newcommand{\pasa}{Publ. Astron. Soc. Aust.} % Publications of the Astron. Soc. of Australia
\newcommand{\pasj}{Publ. Astron. Soc. Jpn} % Publications of the Astron. Soc. of Japan (note no full stop following Jpn)
\newcommand{\pasp}{Publ. Astron. Soc. Pac.} % Publications of the Astron. Soc. of the Pacific
\newcommand{\rmxaa}{Rev. Mexicana Astron. Astrofis.} % Revista Mexicana de Astronomia y Astrofisica
\newcommand{\sci}{Science} % Science
\newcommand{\sciadv}{Sci. Adv.} % Science Advances
\newcommand{\solphys}{Sol. Phys.} % Solar Physics
\newcommand{\sovast}{Soviet Ast.} % Soviet Astronomy
\newcommand{\ssr}{Space Sci. Rev.} % Space Science Reviews
\newcommand{\uni}{Universe} % Universe

\maketitle

\section{Introduction}
The Square Kilometre Array (SKA) will revolutionize the study of star formation in galaxies using deep radio continuum observations \citep{beswick_etal_2015,jarvis_etal_2015,prandoni_and_seymour_2015}. In particular, SKA-Mid telescope observing at 0.35--15.4~GHz with 197 antennas for the SKA baseline design, AA4, enables a clean measurement of the galactic star formation rate (SFR) as they probe higher rest frame frequencies of galaxies with increasing redshift, where emission becomes dominated by thermal (free-free) radiation that is largely extinction free and can be directly related to the ionizing photon rate arising from newly formed massive stars \citep{murphy_etal_2015}.

However, Anomalous Microwave Emission (AME) may appear as excess emission in the rest-frame 10---60~GHz \citep[][]{kogut_etal_1996,leitch_etal_1997,dickinson_etal_2018}, with a typical peak around 30 GHz, although in some cases the peak frequency can reach up to 100 GHz \citep[][]{alihamoud_etal_2009}. Because AME is prominent at similar frequencies where free-free emission is observed (typically rest-frame 33 GHz) this has the potential to contaminate SFR measurement using free-free emission from radio continuum observations at a single observing frequency. SKA-Mid Band 5b will observe the lower frequency side of AME for local galaxies and the peak of the AME SED for $z>2$ galaxies. The prospect of the AME study,(i.e., physical properties and origin of AME) using the SKA has been presented by \cite{dickinson_etal_2015} which also alludes that the integrated contribution of AME from external galaxies could be small in general because AME may only be strongly emitted in a small volume of an entire galaxy. 

Multi-frequency radio observation in more than 3 bands to characterize the AME contribution to SFR measurements is expensive: an important question for extragalactic astronomers who study cosmic SF evolution is how significant the AME contribution to the observed emission at a single frequency is. Since one of the primary goals of SKA is to study the cosmic history of star formation (SF) using radio continuum emission, we need to understand how significant AME emission relative to the thermal free-free emission is for extragalactic star forming regions in the relevant observing parameter spaces (observing frequency and spatial resolution) before we can use the thermal free-free radio emission as SFR tracer and study the star formation in galaxies across the cosmic time in the SKA era. In this chapter, rather than focusing on studying AME itself, we investigate the impact of AME on the observed radio continuum emission from extragalactic star forming regions. 

Although the leading theory of AME emission is electric dipole radiation from nanometer-size spinning dust particles \citep{draine_lazarian_1998, alihamoud_etal_2009}, the carrier of the AME and the physical conditions of the ISM for AME are still not clearly understood \citep[][for review]{dickinson_etal_2018}. For example, PAH were proposed as AME carrier but were not found to correlate with Galactic AME \citep{hensley_etal_2016}, which then suggests other AME carriers such as ultrasmall sillicates, or other emission mechanisms such as magnetic dipole emission \citep{hensley_etal_2016}. Other AME carriers such as amorphous dust \citep{nashimoto_etal_2020} and spinning non-carbon grains \citep{ysard_etal_2022} are also considered. But the observed brightness of AME from various ISM environments does not seem to be explained by a single carrier and therefore in this chapter, we simply assume that AME arises from spinning small particles with an electric dipole and do not attempt to study the nature of AME carriers. 

Unlike AME in the Milky Way where the fraction of AME to the total emission at frequencies near $\approx 30$ GHz for AME detected molecular clouds is as large as 50\% \citep{dickinson_etal_2018}, extragalactic AME is rare: spatially resolved AME is detected from only two galaxies \citep{murphy_etal_2010,scaife_etal_2010,murphy_etal_2018} and galaxy integrated AME is observed from M31 and NGC~2903 only \citep{battistelli_etal_2019,fernandez_torreiro_etal_2024,poojon_etal_2024}. Recently, only upper limits of AME could be inferred for other nearby galaxies (M33, NGC~253, NGC~4945, LMC, and SMC) by accounting for the background CMB fluctuations in these galaxies \citep{correia_etal_2026}. While thermal dust emission arises from large grains that spread over large spatial scales and are heated by the interstellar radiation field, AME is linked to specific grain populations in particular environments. As a result, the localized AME becomes diluted when averaged over entire galaxies \citep{bianchi_etal_2022,correia_etal_2026,dickinson_etal_2018,planck_2014,tibbs_etal_2012}, which highlights AME as a probe of small-scale ISM conditions and the need for high-resolution observations to map its distribution and excitation \citep{correia_2025}.

The physical conditions of the regions in our Milky Way where AME has been detected are diverse: diffuse ionized medium, cold dense molecular clouds, star forming complex, and photo-dissociation regions \citep[e.g.,][]{ade_etal_2011,ade_etal_2014,casassus_etal_2008,watson_etal_2005}. This suggests that the brightness of AME is likely determined by local ISM conditions, for example, density, temperature, interstellar radiation field strength, and ionization fraction as illustrated in theoretical models \citep{alihamoud_etal_2009}, which means that the brightness of the extragalactic AME observed with a finite beam resolution ($0.1-1''$) encompassing a few $\times$ (100 -- 1000) parsec scale depends on the local ISM conditions of the unresolved regions inside the synthesized beam, with different beam filling factors. Also, the thermal free-free emission due to SF aggregated from all regions inside the beam may dominate the observed radio emission. Furthermore, a decrease of the number of small dust particles due to rotational disruption \citep[e.g.,][]{hoang_and_tram_2019} may suppress the AME \citep{yoon_2022}.    

In this chapter, we investigate the impact of AME on the observed radio continuum emission by comparing the thermal free-free emission from extragalactic SF regions with simplified geometry, against an AME model for a range of the observing parameters (observing frequencies, angular resolution, and galaxy redshift) and phases of the ISM. We take a simple analytic approach for estimating the flux density of thermal free-free emission based on star formation rate. The predicted flux density of AME is estimated based on the AME emissivity model \citep{alihamoud_etal_2009}. For simplicity, we will not consider any potential disruption process of small dust particles (i.e., AME carrier) in our investigation. The purpose of this investigation is to benchmark the result against the resolution and sensitivity for SKA-Mid \citep{braun2019anticipatedperformancesquarekilometre} and suggest how best to design the observing strategy for future SKA observations to measure thermal free-free emission for obtaining a reliable SFR measurement. We adopt a flat Lambda CDM cosmology with $H_0 = 70$ km s$^{-1}$ Mpc$^{-1}$, $\Omega_M = 0.3$, and $\Omega_\Lambda = 0.7$. 

\section{Model continuum spectra}
We create model continuum spectra for thermal free-free emission and AME in the frequency range of 1--100 GHz for $z=0.1, 0.3, 1.0, 2.0$. For star forming galaxies, synchrotron emission following a steep power-law spectrum ($\sim \nu^{\alpha}$) with a typical spectral index $\alpha<-0.7$ dominates at lower frequency ($\lesssim 5$ GHz) and rapidly fades at high frequency ($>10$GHz). The radio continuum SED at $\approx 10-30$ GHz is dominated by thermal free-free emission \citep[e.g.,][]{linden_etal_2020,murphy_etal_2012}.

\subsection{Thermal free-free emission}
The ionizing photon production rate determined by the SFR for an assumed IMF is directly proportional to the thermal free-free luminosity $L_{\nu}^{\mbox{\tiny ff}}$. Therefore we can estimate the flux density of thermal free-free emission based on the SFR and gas surface density $\Sigma_{\mbox{\tiny gas}}$. Using the correlation between gas surface density and SFR surface density for a Salpeter IMF, \cite{kennicutt_1998} derived:  
\begin{equation}\label{eq:sfr1}
\frac{\Sigma_{\mbox{\tiny SFR}}}{\mbox{M}_{\odot}\mbox{yr}^{-1}\mbox{kpc}^{-2}} = 2.5\times10^{-4}\left(\frac{\Sigma_{\mbox{\tiny gas}}}{\mbox{M}_{\odot}\mbox{pc}^{-2}}\right)^{1.4}     
\end{equation} Combining this with the following equation in \cite{murphy_etal_2011}, 
\begin{equation}\label{eq:sfr2}
\frac{\mbox{SFR}}{\mbox{M}_{\odot} \mbox{yr}^{-1}} = 4.6\times10^{-28}\left(\frac{T_{\mbox{\tiny e}}}{10^4\mbox{\small K}}\right)^{-0.45}\left(\frac{\nu}{\mbox{GHz}}\right)^{0.1}\left(\frac{L_\nu^{\mbox{\tiny ff}}}{\mbox{erg s}^{-1}\mbox{Hz}^{-1}}\right)
\end{equation} where a Kroupa IMF is used to convert ionizing photon production rate to SFR and $T_{\mbox{\tiny e}}$ and $\nu$ are electron temperature and observing frequency respectively, we obtain 
\begin{equation}
\Sigma_{\mbox{\tiny SFR}}\frac{\pi}{4}\left(\frac{\ell_{\mbox{\tiny ff}}}{\mbox{kpc}}\right)^2 = 3.06\times10^{-28}\left(\frac{T_{\mbox{\tiny e}}}{10^4\mbox{\small K}}\right)^{-0.45}\left(\frac{\nu}{\mbox{GHz}}\right)^{0.1}\left(\frac{L_\nu^{\mbox{\tiny ff}}}{\mbox{erg s}^{-1}\mbox{Hz}^{-1}}\right) 
\end{equation} where $\ell_{\mbox{\tiny ff}}$ is the diameter of the region emitting thermal free-free emission used to measure the SFR. Here we reduced the coefficient in Equation~\ref{eq:sfr2} by factor of 1.5 (from $4.6\times10^{-28}$ to $3.06\times10^{-28}$) to account for the different IMFs used and to render the relation between ionizing photon production rate and SFR used in \cite{murphy_etal_2011} consistent with that in \cite{kennicutt_1998}. If we assume a cylindrical geometry and adopt the relation between neutral hydrogen column density of the free-free emission region $N^{\mbox{\tiny ff}}_{\mbox{\tiny H}}$ and $\Sigma_{\mbox{\tiny gas}}$ using Equation (4) in \cite{rahmati_etal_2013} with a 75\% hydrogen mass fraction
\begin{equation}
N^{\mbox{\tiny ff}}_{\mbox{\tiny H}}\approx10^{20}\mbox{cm}^{-2}\left(\frac{\Sigma_{\mbox{\tiny gas}}}{\mbox{M}_{\odot}\mbox{pc}^{-2}}\right),
\end{equation} we can express the thermal free-free luminosity $L_{\nu}$ as follows
\begin{equation}
\left(\frac{L_\nu^{\mbox{\tiny ff}}}{\mbox{erg s}^{-1}\mbox{Hz}^{-1}}\right) = 6.39\times10^{23}\left(\frac{\ell_{\mbox{\tiny ff}}}{\mbox{kpc}}\right)^2\left(\frac{N^{\mbox{\tiny ff}}_{\mbox{\tiny H}}}{10^{20}\mbox{cm}^{-2}}\right)^{1.4}\left(\frac{T_{\mbox{\tiny e}}}{10^4\mbox{\small K}}\right)^{0.45}\left(\frac{\nu}{\mbox{GHz}}\right)^{-0.1}
\end{equation}

Finally, the free-free emission flux density $S_\nu^{\mbox{\tiny ff}}$ for a source at luminosity distance $D_{\mbox{\tiny L}}$ is
\begin{equation}\label{eq:ff_flux1}
\left(\frac{S_\nu^{\mbox{\tiny ff}}}{\mu\mbox{Jy}}\right) = 537\times\left(\frac{D_{\mbox{\tiny L}}}{\mbox{Mpc}}\right)^{-2}\left(\frac{\ell_{\mbox{\tiny ff}}}{\mbox{kpc}}\right)^2\left(\frac{N^{\mbox{\tiny ff}}_{\mbox{\tiny H}}}{10^{20}\mbox{cm}^{-2}}\right)^{1.4}\left(\frac{T_{\mbox{\tiny e}}}{10^4\mbox{\small K}}\right)^{0.45}\left(\frac{\nu}{\mbox{GHz}}\right)^{-0.1}
\end{equation}

For our convenience, we replace $\ell_{\mbox{\tiny ff}}$ with the observing beam size $\theta$ by assuming that the free-free emission region is large enough to fill the observing beam and use angular diameter distance $D_{\mbox{\tiny A}}$ instead of luminosity distance $D_{\mbox{\tiny L}}$ for source redshift $z$. Using $D_{\mbox{\tiny A}}\theta=\ell_{\mbox{\tiny ff}}$ and $D_{\mbox{\tiny L}}=(1+z)^2D_{\mbox{\tiny A}}$, we obtain the following expression for $S_{\nu}^{\mbox{\tiny ff}}$
\begin{equation}\label{eq:ff_flux2}
\left(\frac{S_\nu^{\mbox{\tiny ff}}}{\mu\mbox{Jy}}\right) = 12.62\times10^{-3}\frac{1}{(1+z)^4}\left(\frac{\theta}{1''}\right)^{2}\left(\frac{N^{\mbox{\tiny ff}}_{\mbox{\tiny H}}}{10^{20}\mbox{cm}^{-2}}\right)^{1.4}\left(\frac{T_{\mbox{\tiny e}}}{10^4\mbox{\small K}}\right)^{0.45}\left(\frac{\nu}{\mbox{GHz}}\right)^{-0.1}
\end{equation}

\subsection{Anomalous Microwave Emission}
We use IDL software \texttt{SpDust} \citep{alihamoud_etal_2009} to compute the emissivity of AME per hydrogen atom, $j_{\nu}/n_{\mbox{\tiny H}}$ (Jy sr$^{-1}$cm$^2$ per H-atom). In \texttt{SpDust}, the input parameters are the intrinsic ISM properties and therefore the observational bias such as HI saturation in high density cannot be properly considered in \texttt{SpDust}. For a source at luminosity distance $D_{\mbox{\tiny L}}$, the observed flux density of AME, $S_\nu^{\mbox{\tiny AME}}$, by assuming a cylindrical geometry for the AME emitting region with the diameter $\ell_{\mbox{\tiny AME}}$ and the size $l$ along the line-of-sight becomes 
\begin{equation}
    S_\nu^{\mbox{\tiny AME}} = \frac{\pi}{4}\ell_{\mbox{\tiny AME}}^2 n_{\mbox{\tiny H}} \left(\frac{4\pi j_\nu}{n_{\mbox{\tiny H}}}\right)l \bigg/ 4\pi D_{\mbox{\tiny L}}^2
\end{equation} where $n_{\mbox{\tiny H}}$ is the hydrogen number density (cm$^{-3}$). In terms of observed units, we can rewrite it as follows 
\begin{equation}\label{eq:ame_flux1}
    \left(\frac{S_\nu^{\mbox{\tiny AME}}}{\mu\mbox{Jy}}\right) = 0.785\times\left(\frac{D_{\mbox{\tiny L}}}{\mbox{Mpc}}\right)^{-2}\left(\frac{\ell_{\mbox{\tiny AME}}}{\mbox{kpc}}\right)^2\left(\frac{N^{\mbox{\tiny AME}}_{\mbox{\tiny H}}}{1\mbox{cm}^{-2}}\right) \left(\frac{j_\nu}{n_{\mbox{\tiny H}}}\right)
\end{equation}
where $N^{\mbox{\tiny AME}}_{\mbox{\tiny H}}$ is the hydrogen column density of an AME emitting region (i.e.,  $n_{\mbox{\tiny H}} l$)

Likewise, using an observing beam size $\theta$, angular diameter distance $D_{\mbox{\tiny A}}$ ($D_{\mbox{\tiny A}}\theta=\ell_{\mbox{\tiny AME}}$), $S_\nu^{\mbox{\tiny AME}}$ can be rewritten as 
\begin{equation}\label{eq:ame_flux2}
    \left(\frac{S_\nu^{\mbox{\tiny AME}}}{\mu\mbox{Jy}}\right) = 1.84\times10^{-5}\frac{1}{(1+z)^4} \left(\frac{\theta}{1''}\right)^2 \left(\frac{N^{\mbox{\tiny AME}}_{\mbox{\tiny H}}}{1\mbox{cm}^{-2}}\right) \left(\frac{j_\nu}{n_{\mbox{\tiny H}}}\right)
\end{equation}
However, we note that the angular size of the AME region ($\theta_{\mbox{\tiny AME}}=\ell_{\mbox{\tiny AME}}/D_{\mbox{\tiny A}}$) may or may not fill the observing beam ($\theta_{\mbox{\tiny AME}}<\theta$) unlike thermal free-free emission is at a larger scale. As we will present in Section 3, one of the main goals of this chapter is to investigate the relative fraction of AME in the observed radio flux density at a given observing frequency for different sizes of the observing beam from 0.1 to 1$^{''}$ and for a fixed size of AME emitting region from a source at different redshifts by assuming that free-free emission from the star-forming region fills the observing beam while AME from a smaller scale may not. Therefore when we create the radio SED of an AME, we will use Equation~\ref{eq:ame_flux1} by assuming the diameter of AME emitting region ($\ell_{\mbox{\tiny AME}}$).

One of the key parameters that changes the strength of AME emissivity is the energy density of ambient radiation field $\chi$ in units of the local interstellar radiation field energy density $u_{\mbox{\tiny ISRF}}$. The original version of \texttt{SpDust} is modified to include the CMB as the ambient radiation field. At high-$z$, the ambient radiation field intensity increases with the increasing CMB energy density. Therefore the new ambient radiation field energy density $\chi'$ at redshift $z$ is related to the \texttt{SpDust} input parameter for the radiation field energy density $\chi$ at $z=0$ as follows.
\begin{equation}
    \chi' u_{\mbox{\tiny ISRF}} = \chi u_{\mbox{\tiny ISRF}} + U_{\mbox{\tiny CMB}}(z)
\end{equation}
Therefore,
\begin{equation}
    \chi' = \chi + \frac{U_{\mbox{\tiny CMB}}(0)}{u_{\mbox{\tiny ISRF}}}(1+z)^4
\end{equation} which implies that the AME emissivity from star-forming regions in galaxies will increase at high-$z$. We adopted $U_{\mbox{\tiny CMB}}(0)=4.2\times10^{-13}$ erg cm$^{-3}$ and $u_{\mbox{\tiny ISRF}}=1.05\times10^{-12}$ erg cm$^{-3}$ \citep{draine_2011}, and modified the original \texttt{SpDust} code creating an adjusted ambient radiation field by adding a CMB contribution and used $\chi'$ instead of $\chi$ in \texttt{SpDust} when we compute the AME emissivity at redshift $z$.

\subsection{Composite radio SED from free-free emission and AME}
We create a composite radio continuum SED from a combination of thermal free-free emission using Equation~\ref{eq:ff_flux2} and AME using Equation~\ref{eq:ame_flux1}. We assume that thermal free-free emission is from the warm ionized medium with a fixed electron kinetic temperature $T_{\mbox{\tiny e}}=10^4$K. For AME, we create the emissivities for four different ISM phases: cold neutral medium (CNM), warm neutral medium (WNM), warm ionized medium (WIM), and photo-dissociation region (PDR), using the representative physical parameters from \cite{draine_lazarian_1998}. 

\begin{figure}[h]
    \centering
	\includegraphics[width=0.48\columnwidth]{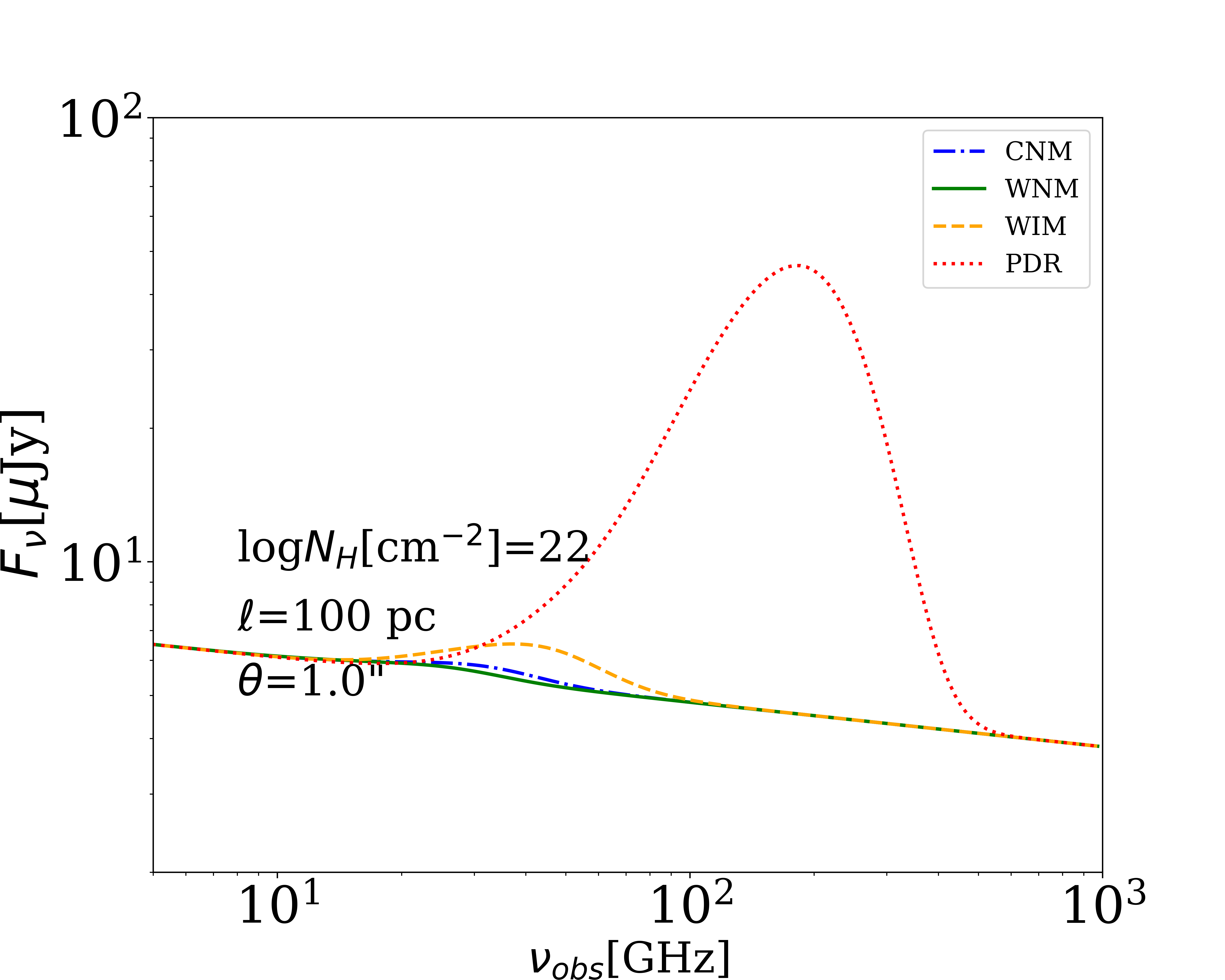}
    \includegraphics[width=0.48\columnwidth]{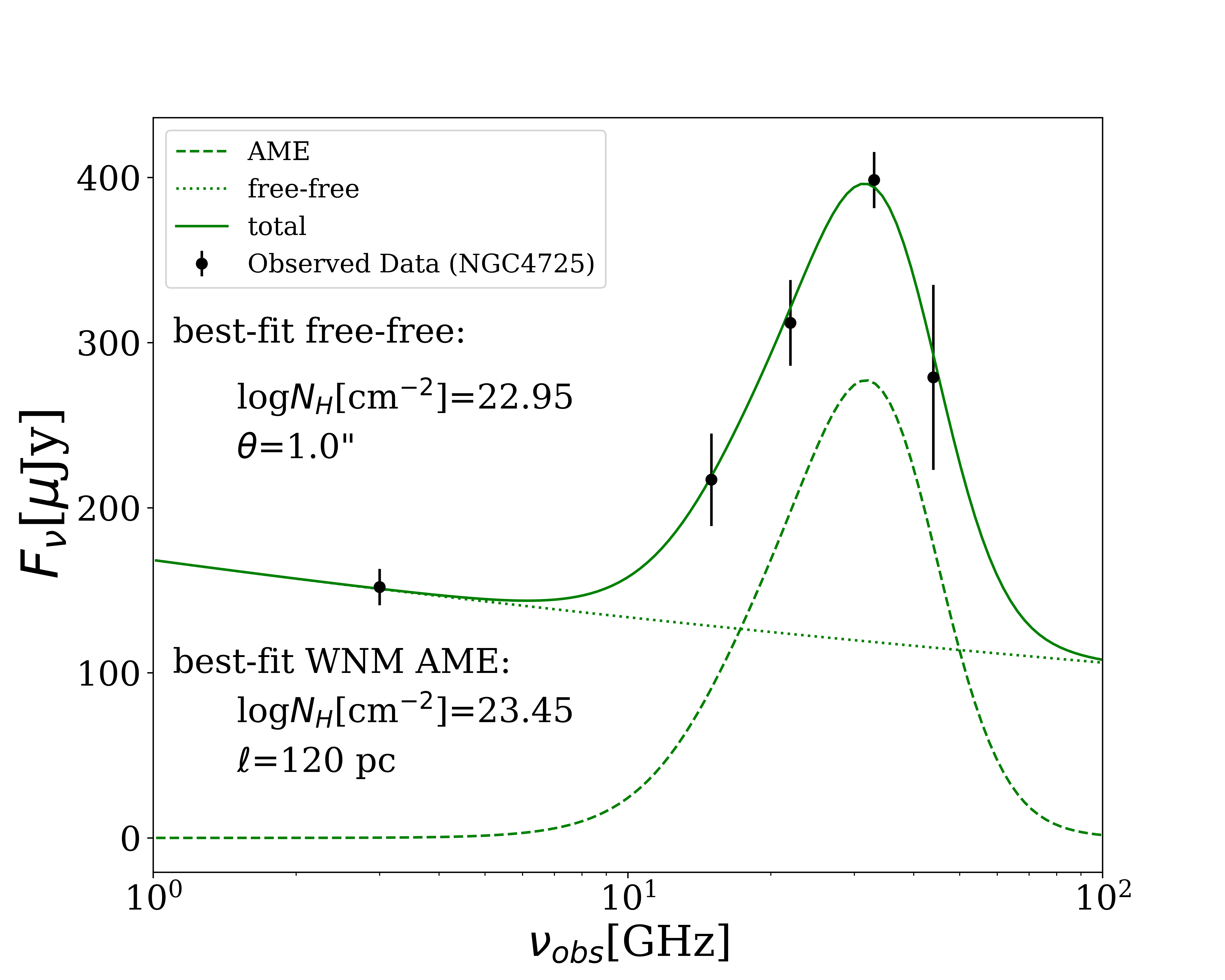}
    \caption{\textit{Left: }Model radio continuum SED for a galaxy at $z=0.001$ using different phases of the ISM for the AME. \textit{Right:} Observed radio continuum emission (black dots with error bars) from the AME region in NGC~4725 \citep{murphy_etal_2018} and the best-fit model SED using free-free emission and AME.}
    \label{fig:example}
\end{figure}

In Figure~\ref{fig:example}, we present two examples of our model radio SED. In the left panel of Figure~\ref{fig:example}, we show a model radio SED for a fiducial nearby galaxy at $z=0.001$ using thermal free-free emission filling a 1$''$ beam (22pc at $z=0.001$) and AME for different phases of the ISM with a diameter of 100 pc. For the same column density for both free-free emission and AME ($N^{\mbox{\tiny ff}}_{\mbox{\tiny H}}=N^{\mbox{\tiny AME}}_{\mbox{\tiny H}}=10^{22}$cm$^{-2}$), the AME flux density varies significantly depending on the ISM phase. AME from a PDR is the strongest. However, the actual PDR size is small (a few parsec) and the PDR volume filling factor should be much smaller than the other phases of a ISM. In the left panel of Figure~\ref{fig:example}, if we assume that the PDR size is 10\% of the CNM/WNM/WIM ($\approx 10$ pc), the AME flux density from PDR is decreased by 1/100 (see Equation~\ref{eq:ame_flux1}) and becomes negligible compared to the AME from other phases of the ISM. Therefore it is reasonable to assume that the contribution of PDR to AME should be minor unless the observing telescope has enough angular resolution and sensitivity to detect the individual PDR associated with the extragalactic star-forming region. The most relevant ISM phases for our investigation would be WNM or WIM. 

In the right panel of Figure~\ref{fig:example}, we present the model radio continuum SED from free-free emission and AME by adopting $\theta=1''$ observing beam and using the value of $N_{\mbox{\tiny H}}$ for each component and the AME emitting region diameter $\ell_{\mbox{\tiny AME}}$. The model SED fits the measured radio continuum data (black dots) from a localized region in NGC~4725 based on the multi-band VLA observation by \cite{murphy_etal_2018}. The high column density values ($N_{\mbox{\tiny H}}\approx 10^{23}$cm$^{2}$) and the assumed observing beam size ($\theta=1''$) used to fit the observed data implies that the AME in NGC~4725 is from a localized high-density environment with $\ell_{\mbox{\tiny AME}}=120$ pc, which is consistent with the inferred column density ($\approx 10^{23}$cm$^{2}$) and the angular size of the AME emission (1--2$''$ or 60--120 pc at 11.9Mpc distance of the source) in \cite{murphy_etal_2018}.

We note that WNM may not have $N_{\mbox{\tiny H}}\approx 10^{23}$cm$^{-2}$. Such high column density may lead to an overprediction of AME and an increased optical depth of free-free emission that becomes optical thick typically at $\lesssim 1$ GHz. However, given the inability of \texttt{SpDust} for considering HI saturation mentioned above, our calculation can be taken as an upper limit of AME and the subsequent results presented in Section 3 will demonstrate that even if we push the ISM condition to the extremely high $N_{\mbox{\tiny H}}$ such that it may overpredict the AME brightness, the AME contribution in the observed radio emission is not significant compared to the free-free emission. Also, the free-free optical depth does not only depend on emission measure (EM) but also on the rest frequency ($\tau_{\mbox{\tiny ff}}\sim T_e^{-1.35}\nu^{-2.1}$EM). Therefore, at 10GHz frequency with $\tau_{\mbox{\tiny ff}}$ being $>100$ times lower than at 1 GHz for the same EM, the free-free emission is likely to be optically thin. Figure~\ref{fig:example} demonstrates that the proposed radio continuum SED model matches the observation reasonably well and justifies the validity of our approach to investigate the impact of AME on the radio continuum SED. 

\section{Result}
For our investigation, we assume that free-free emission and AME originate from regions with the same column density ($N^{\mbox{\tiny ff}}_{\mbox{\tiny H}}=N^{\mbox{\tiny AME}}_{\mbox{\tiny H}}=N_{\mbox{\tiny H}}$) and the free-free emission emitting region fills the observing beam while the AME emitting region may not. We will present the ensemble of the model radio continuum SEDs for different hydrogen column density $N_{\mbox{\tiny H}}$ and the AME region size (henceforth, referred to $\ell$ for $\ell_{\mbox{\tiny AME}}$), observing beam size $\theta$, and redshift $z$, for distant and nearby galaxies. The model SED will be compared against the resolution and sensitivity in Band 5b (8.3--15.4 GHz) for the SKA-Mid AA4 baseline design. The range of angular resolution of the SKA-Mid AA4 baseline in Band 5b is 0.08-10$''$ (Table 7 in \cite{braun2019anticipatedperformancesquarekilometre}). In this study, we use the SKA sensitivity calculator\footnote{\url{https://sensitivity-calculator.skao.int/}} to estimate the $3\sigma$ continuum sensitivity with 10 hour integration and maximum available bandwidth for three different angular resolutions: approximately 0.1, 0.3, and 1.0$''$ by changing the Briggs' weighting parameter and tapering scale, which results in approximately the same 3 $\sigma$ continuum sensitivity ($\sim 1 \mu$Jy). First, we will present the implications of our investigation for observations using the SKA-Mid AA4 baseline in Section~\ref{sec:result1}. Then we will discuss the flux density above the detection limit and the AME fraction in further detail for different $N^{\mbox{\tiny AME}}_{\mbox{\tiny H}}$, $\ell_{\mbox{\tiny AME}}$, and observing beam size, in Section~\ref{sec:result2} and \ref{sec:result3}.

\begin{figure}[h]
    \centering
	\includegraphics[width=0.48\columnwidth]{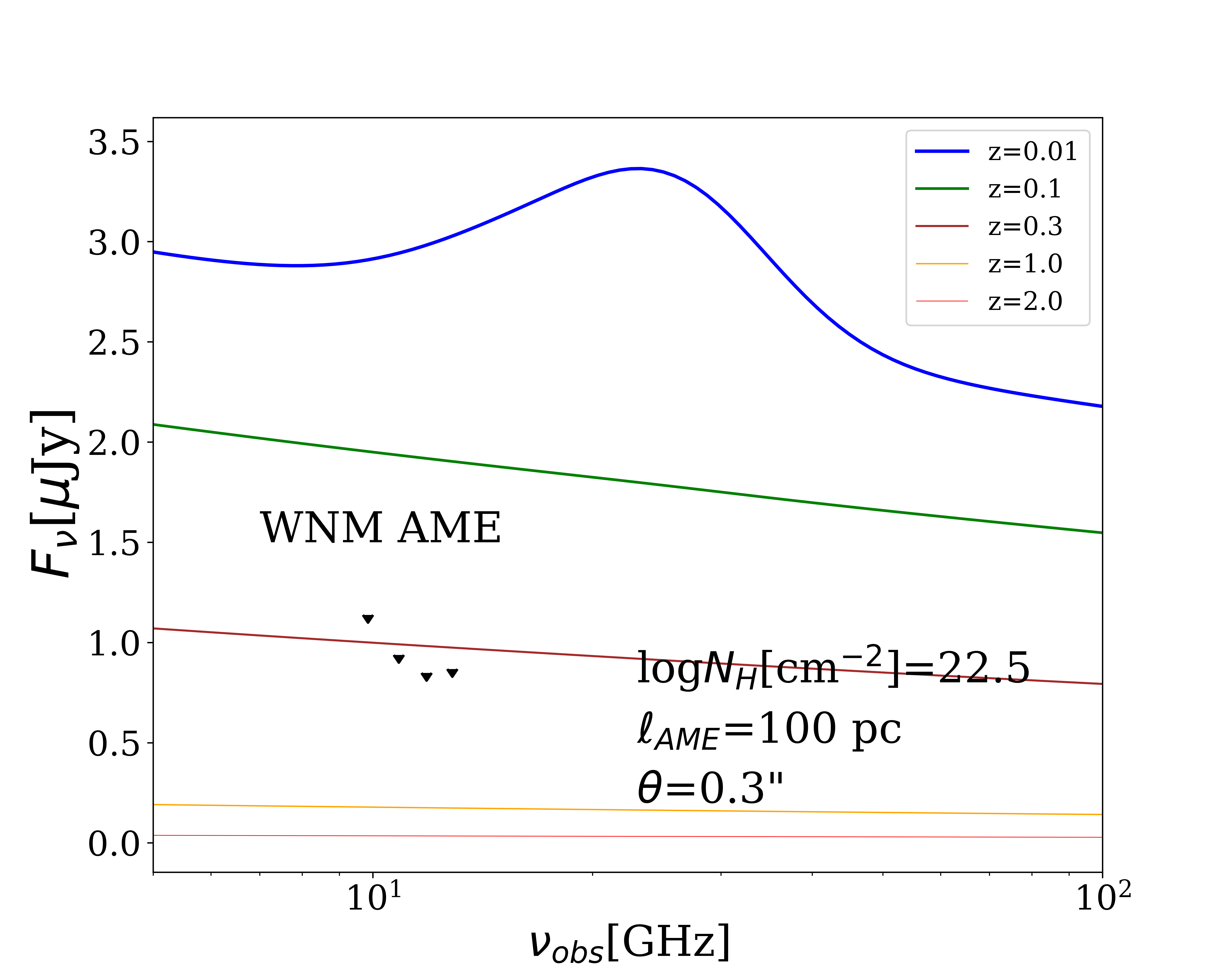}
    \includegraphics[width=0.48\columnwidth]{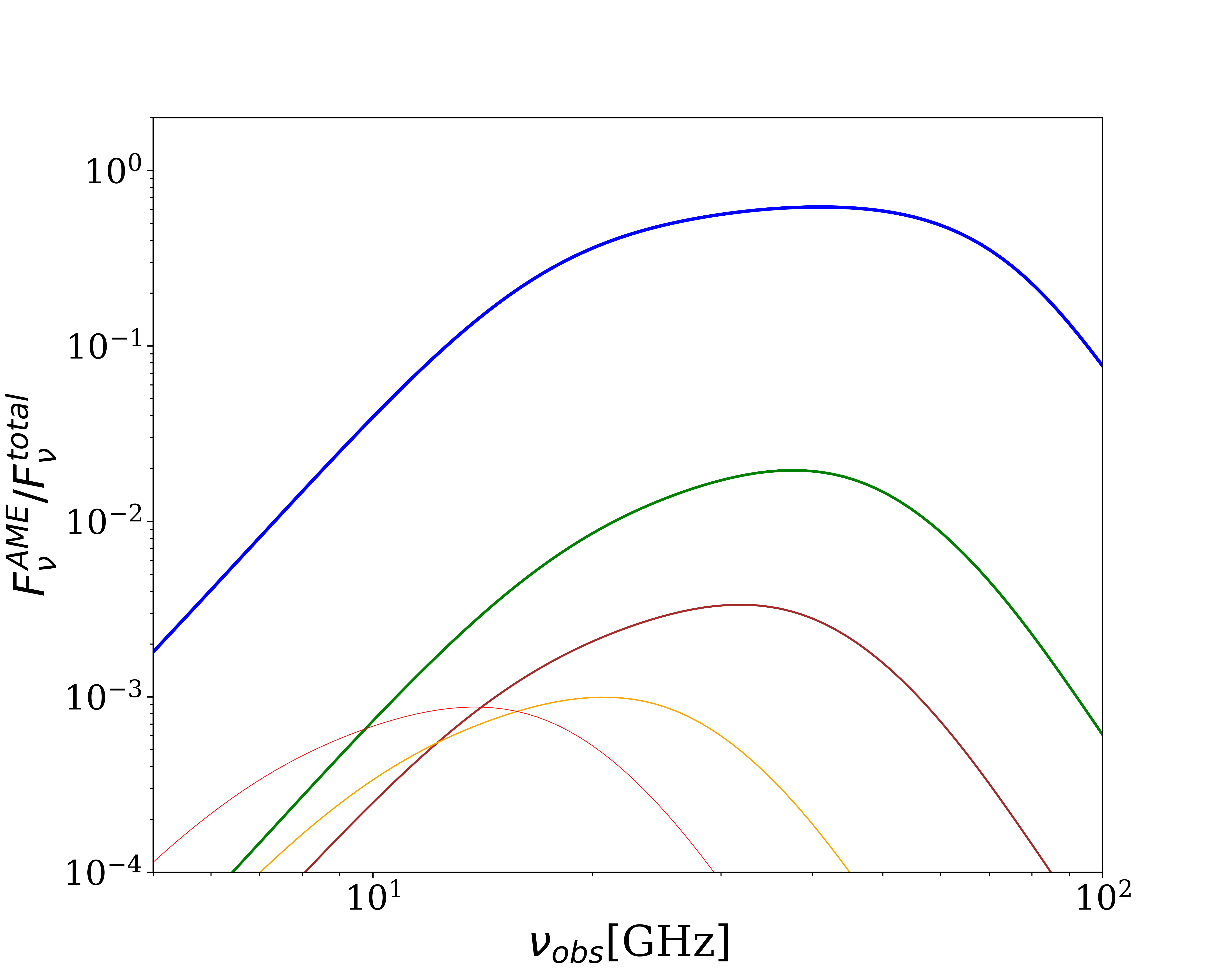}
    \caption{\textit{Left: }Model SED observed with an 0.3$''$ beam for $z=0.01,0.1,0.3,1.0,2.0$ assuming AME from a WNM with $N_{\mbox{\tiny H}}=10^{22.5}$cm$^{-2}$ and $\ell_{\mbox{\tiny AME}}=100$pc, and 3$\sigma$ RMS sensitivities of the Band 5b frequencies with 10 hours of integration shown by black down arrows.   
    \textit{Right:} AME fraction for the model SEDs in the left panel.}
    \label{fig:example_investigation}
\end{figure}

\subsection{Implications for observations with the SKA-Mid}\label{sec:result1}
In Figure~\ref{fig:example_investigation}, we show the radio continuum SED using AME from a WNM with $N_{\mbox{\tiny H}}=10^{22.5}$cm$^{-2}$ and $\ell=100$pc observed by 0.3$''$ beam (left) and the fraction of AME flux density in the total flux density (right), for different redshift ($z=0.01, 0.1, 0.3, 1.0, 2.0$). Down arrows in the left-hand panel indicates the $3\sigma$ RMS continuum sensitivity of four different observing frequencies in Band 5b (9.85, 10.85, 11.85 and 12.85 GHz) for 10 hours of integration. If $z>0.1$, the model flux densities are below $3\sigma$ sensitivity detection limits (left panel) and the AME fraction is less than 1\% (right panel).  

In general, both AME and free-free emission is difficult to detect with increasing redshift. However, given the observing beam size, a relative portion of the AME and the free-free emission in the observed radio continuum emission may vary depending on the physical condition and beam filling factor of the ISM component responsible for each emission process. From $z=0.01$ to 2.0 ($z=0.01$, 0.1, 0.3, 1.0, 2.0), the corresponding angular scales are 20--850 pc for 0.1$^{''}$, 70--2500 pc for 0.3$^{''}$, and 205–-8500 pc for 1.0$^{''}$ beam. For $z\gtrsim0.1$ sources, a spatial scale smaller than 180 pc is not resolved by the highest angular resolution beam (0.1$^{''}$). Therefore, for most cases, the assumed AME region (100 pc) is not resolved by the observing beam we consider, while star-forming complex region is still large enough to fill the beam.

For distant galaxies, one can naively expect that AME is not very bright because of the small beam filling factor of AME in addition to redshift dimming. For nearby galaxies, AME may be a significant fraction of the observed radio emission depending on the relative size of AME region to the observing beam size. Therefore, we investigate the impact of AME separately for distant and nearby galaxies in the following sections, by using typical parameters over a small range. 

\begin{figure}[h]
    \centering
	\includegraphics[width=0.47\columnwidth]{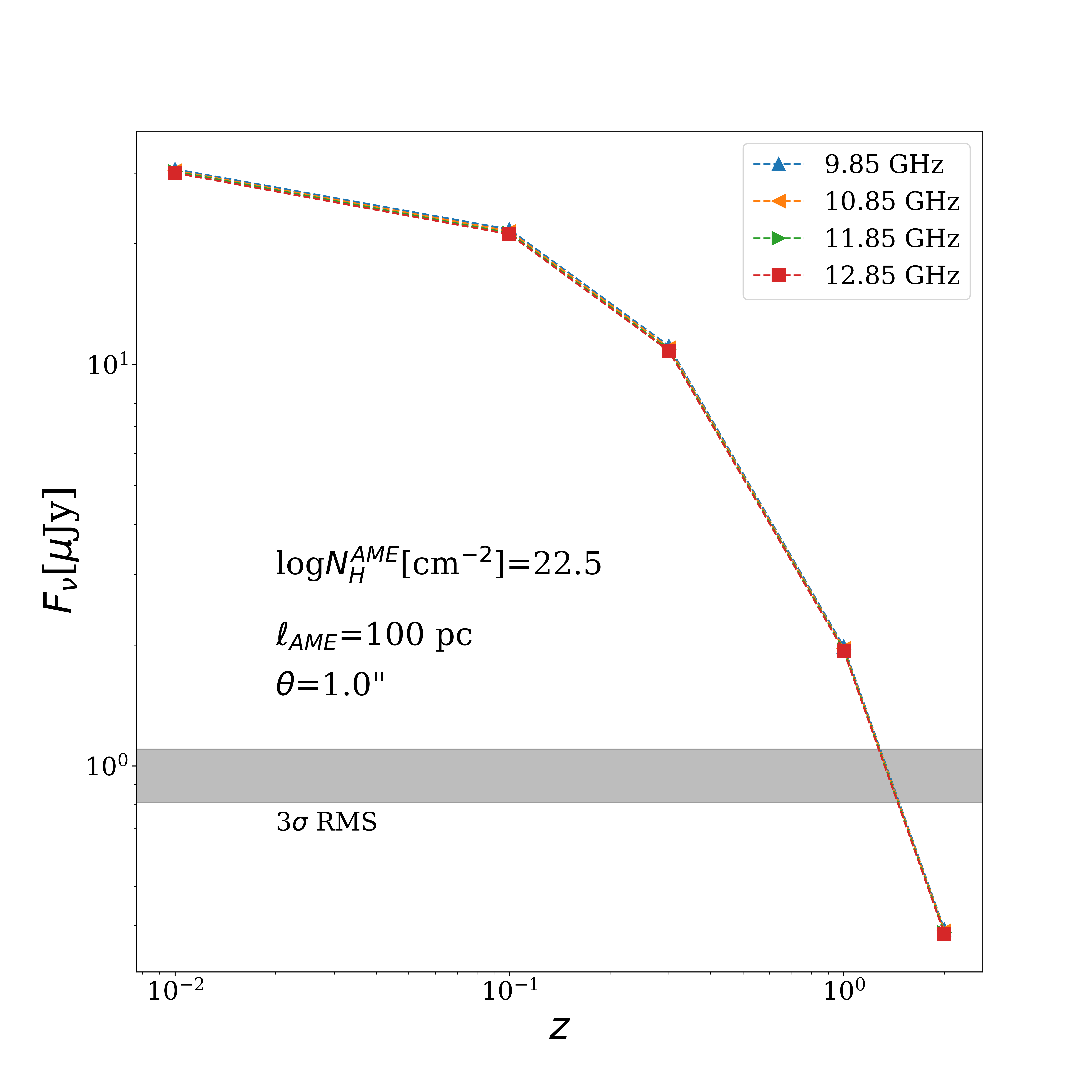}
    \includegraphics[width=0.47\columnwidth]{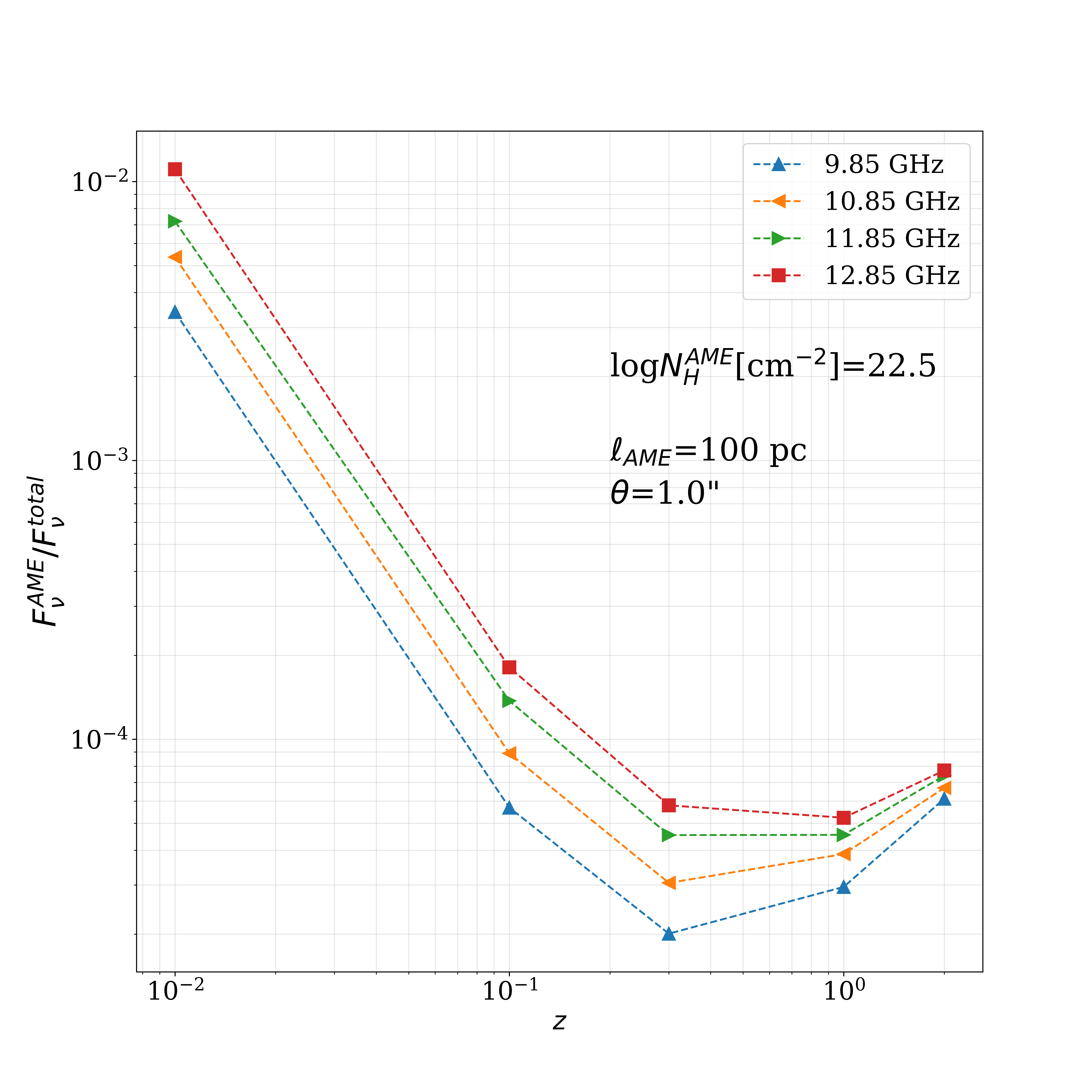}
    \caption{\textit{Left: }Expected flux density of the model SED observed by 1$''$ beam at the observing frequencies (shown by different colors and symbols) of SKA-Mid Band 5b (9.85, 10.85, 11.85, and 12.85 GHz) as a function of redshift for $N_{\mbox{\tiny H}}=10^{22.5}$cm$^{-2}$cm, $\ell=100$pc, and the gray shaded region showing the range of 3$\sigma$ RMS values with 10 hours integration for the chosen observing frequencies.  
    \textit{Right:} AME fraction for the same model SEDs used for the left panel}
    \label{fig:investigation1}
\end{figure}

\subsection{Integrated observations of distant galaxies}\label{sec:result2}
In the left panel of Figure~\ref{fig:investigation1}, we show the expected flux density of the model SED at the observing frequencies (shown by different colors and symbols) of Band 5b as a function of redshift. We consider a galaxy in 5 different redshifts: $z=0.01,0.1,0.3,1.0,2.0$. The luminosity distance for the lowest redshift (43 Mpc at $z=0.01$) is larger than the distances of the two galaxies within which AME is detected: NGC~4725 at 11.9 Mpc \citep{murphy_etal_2018} and NGC~6946 at 6.8Mpc \citep{murphy_etal_2015}. The observing beam size of 1.0$''$ corresponds to 205 pc at $z=0.01$ and $\gtrsim 2$kpc for $z>0.1$. In this section where we consider the integrated properties of galaxies we do not consider beams smaller than $\theta<1''$. An $N_{\mbox{\tiny H}}=10^{22.5}$cm$^{-2}$ ($\approx 300$M$_{\odot}$ pc$^{-2}$) corresponds to gas surface density similar to that of NGC~6946 \citep{crosthwaite_etal_2007} for both free-free emission and AME. The brightness of model SEDs considered in our investigation falls below $3\sigma$ RMS (gray shaded region) if $z>1$. 

The right panel of Figure~\ref{fig:investigation1} shows the fraction of AME in the total radio continuum emission. For all redshifts and frequencies, the AME fraction is $\lesssim 1$\%, which means that the most of the emission is thermal and AME contribution to the radio SED is negligible. If the column density increases, the expected flux density of free-free emission increases proportionally to $N_{\mbox{\tiny H}}^{1.4}$ (Equation~\ref{eq:ff_flux1}) while the expected AME density increases proportionally to $N_{\mbox{\tiny H}}$ (Equation~\ref{eq:ame_flux1}). Therefore the AME contribution becomes even smaller with increasing column density. Overall, the AME has negligible impact on the radio SED for distant galaxies unless the observing beam becomes small and starts to resolve the star forming regions as we discuss for nearby galaxies.

\subsection{Spatially resolved star-forming regions in nearby galaxies}\label{sec:result3}
The SKA can resolve individual extragalactic star-forming regions. Therefore the relative size of the AME emitting region ($\ell_{\mbox{\tiny AME}}$) to the observing beam ($\theta$) becomes important parameter for our investigation. For a 100 pc diameter AME emitting region ($\ell_{\mbox{\tiny AME}}=$100 pc), we calculate the radio SED and the AME fraction for different observing beam sizes ($\theta=0.1, 0.3, 1.0''$).   

\begin{figure}[h]
    \centering
	\includegraphics[width=0.32\columnwidth]{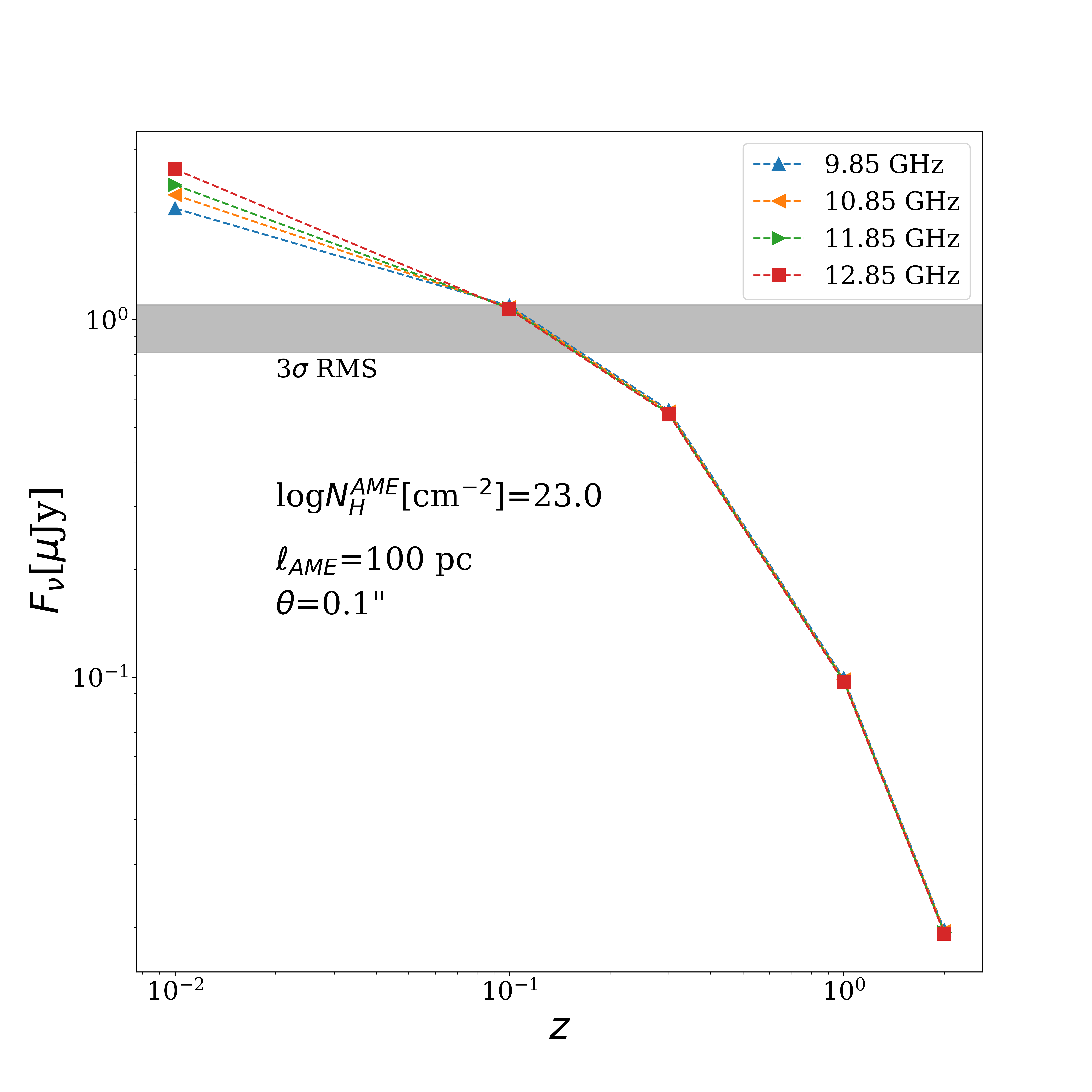}
    \includegraphics[width=0.32\columnwidth]{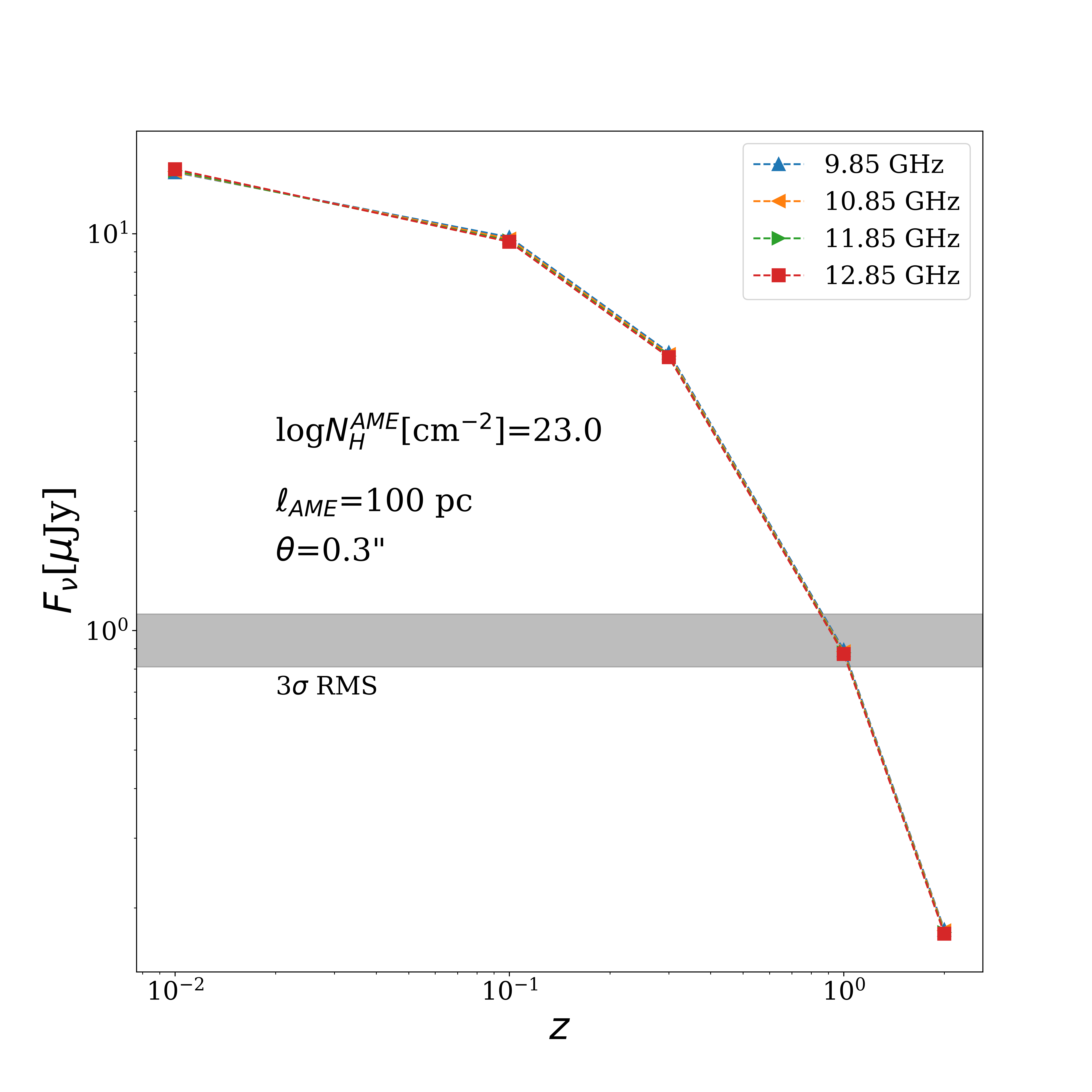}
    \includegraphics[width=0.32\columnwidth]{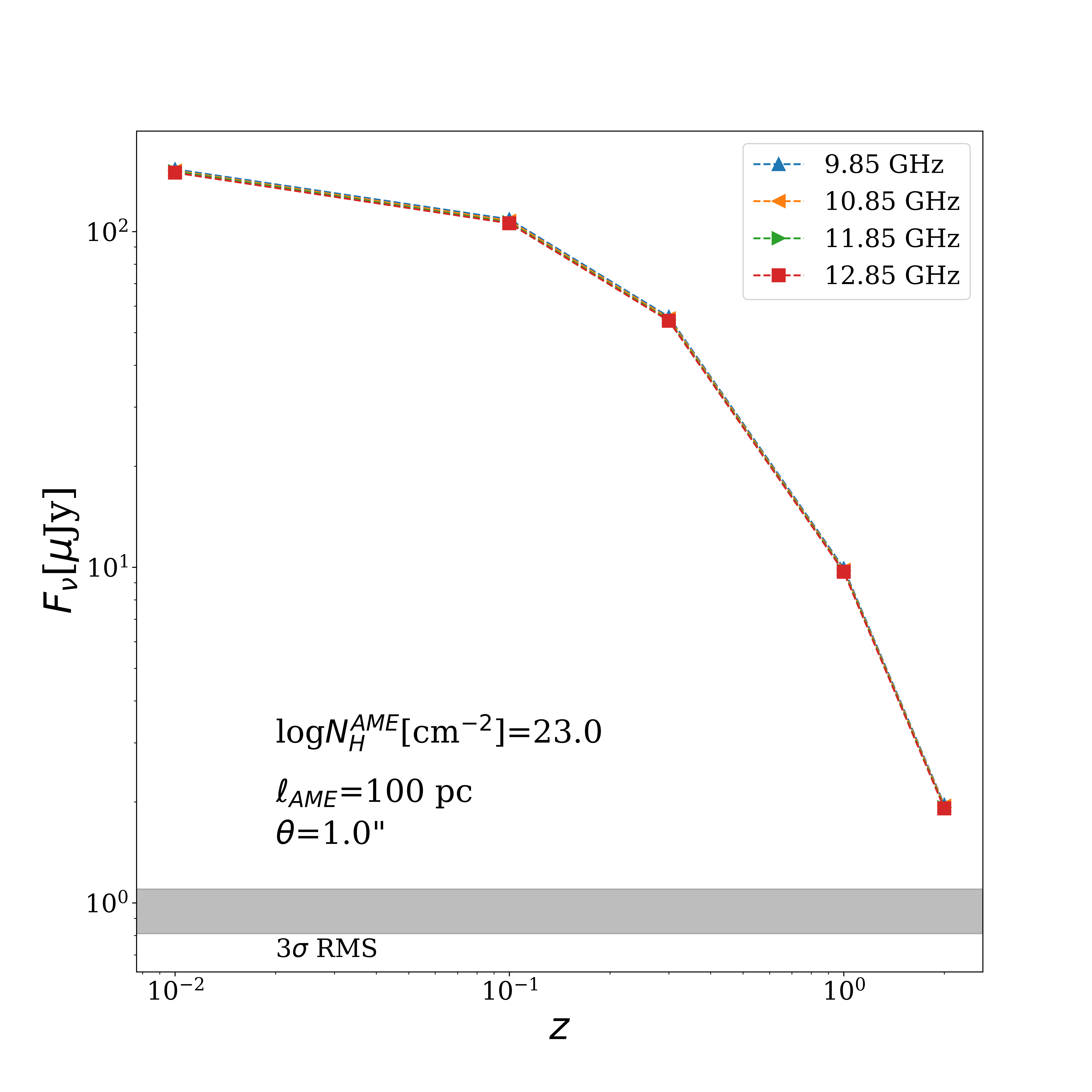}
    \caption{Expected flux density of the model SED observed with a 0.1 (left), 0.3 (middle), and, 1.0$''$(right) beam at the observing frequencies (shown by different colors and symbols) of Band 5b (9.85, 10.85, 11.85, and 12.85 GHz) as a function of redshift for $N_{\mbox{\tiny H}}=10^{23.0}$cm$^{-2}$, $\ell=100$pc. The gray shaded region shows the range of 3$\sigma$ RMS values with 10 hours integration for the chosen observing frequencies.}
    \label{fig:investigation2}
\end{figure}

Figure~\ref{fig:investigation2} shows the expected flux density of the model SEDs at the observing frequencies (shown by different colors and symbols) of SKA-Mid Band 5b (9.85, 10.85, 11.85, and 12.85 GHz) as a function of redshift for $N_{\mbox{\tiny H}}=10^{23.0}$cm$^{-2}$cm and $\ell=100$pc, observed by three different beams: 0.1, 0.3, and 1.0$''$ shown in the left, middle and right panel, respectively. The flux density is $\approx 3\times$ higher than that of the model SED in Figure~\ref{fig:investigation1} due to the increase in column density by the same amount. We note that there is a noticeable difference in the flux density depending on observing frequency when the relative size of an AME region becomes large compared to the observing beam (100pc AME region observed by 0.1$''$ beam shown on the left panel): the flux density is larger for higher observing frequencies because the contribution of AME to the total radio emission becomes significant.

\begin{figure}[h]
    \centering
	\includegraphics[width=0.32\columnwidth]{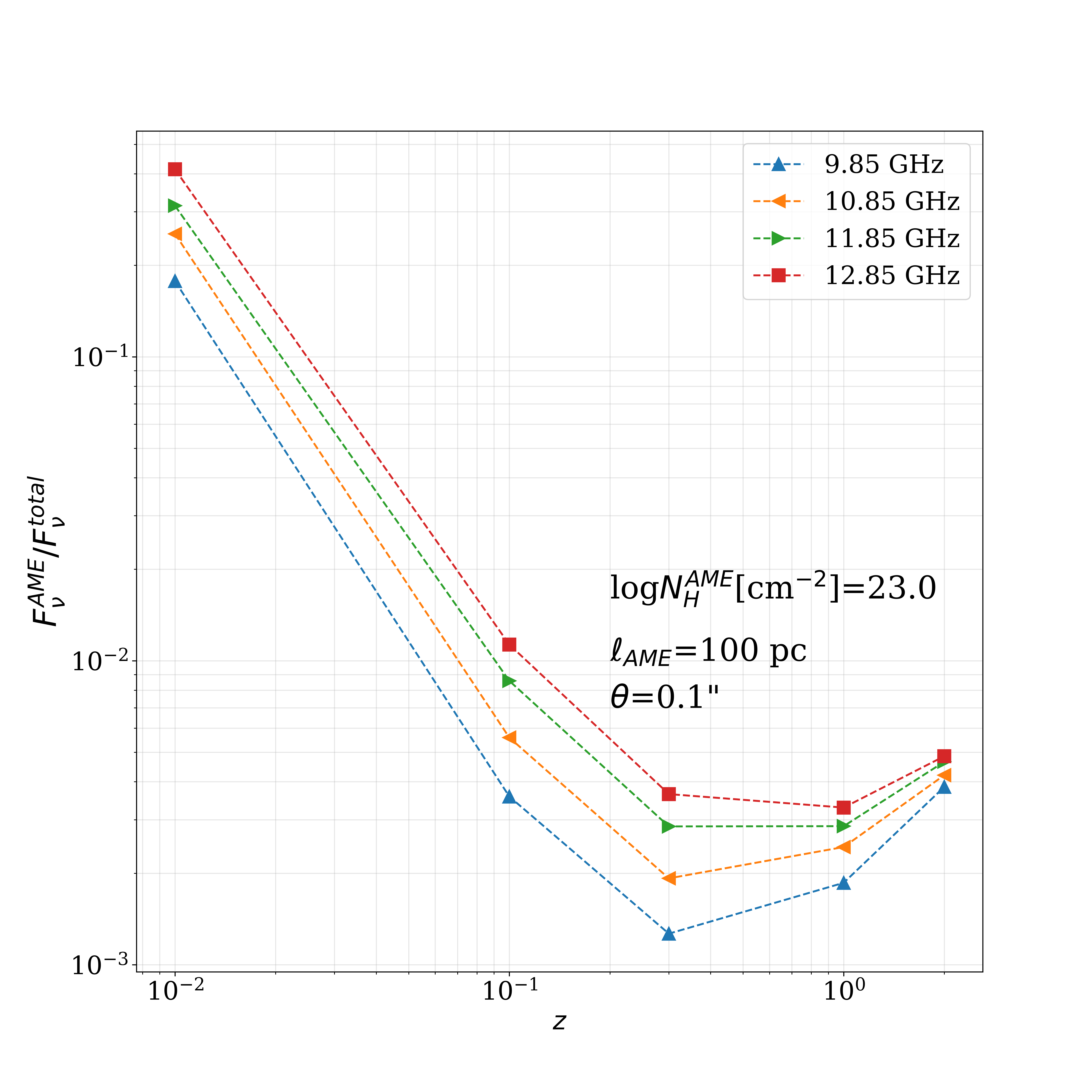}
    \includegraphics[width=0.32\columnwidth]{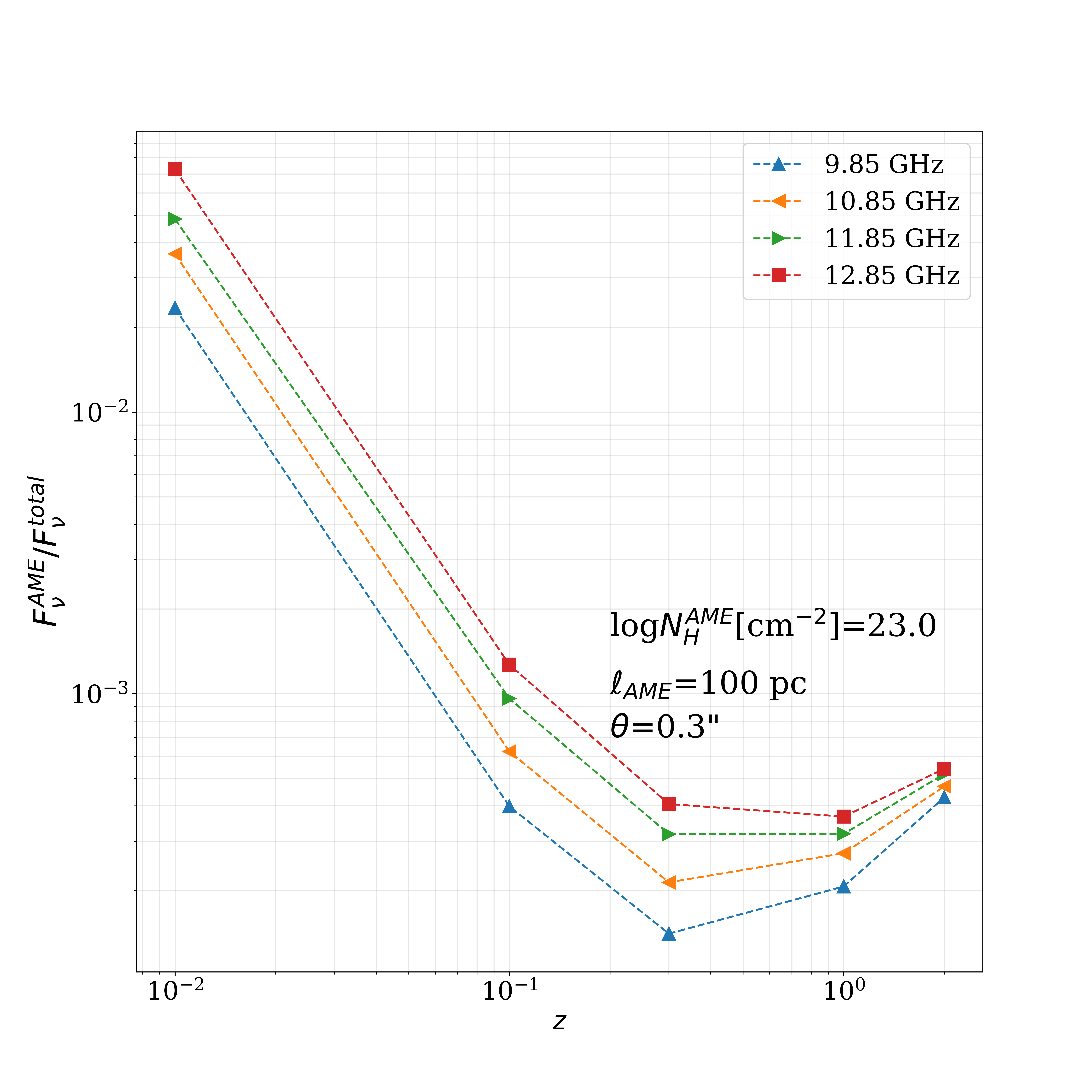}
    \includegraphics[width=0.32\columnwidth]{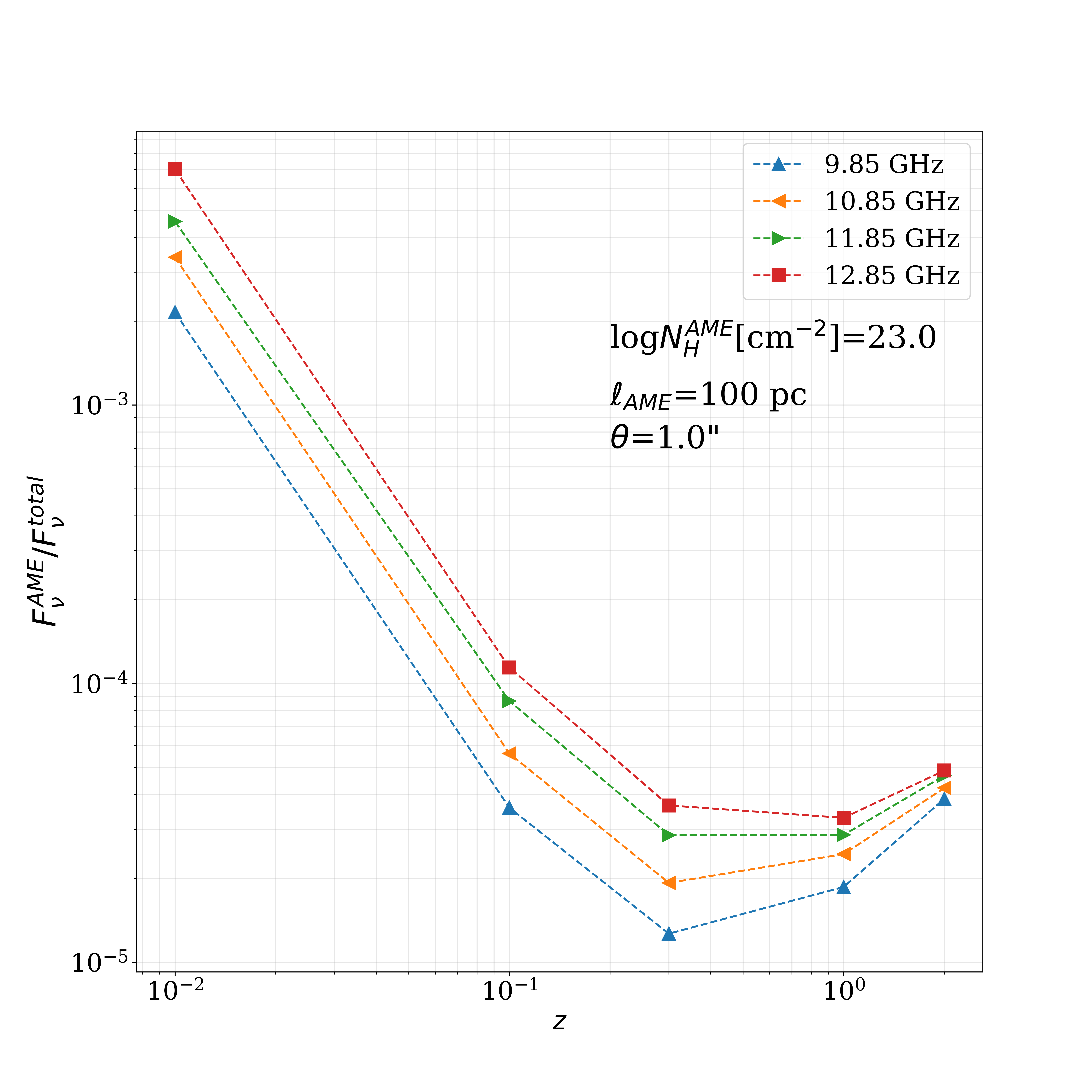}
    \caption{AME fraction for the same model SEDs used for Figure~\ref{fig:investigation2}}
    \label{fig:investigation3}
\end{figure}

In Figure~\ref{fig:investigation3}, each panel shows the fraction of AME to the total radio continuum emission corresponding to the matched panels in Figure~\ref{fig:investigation2}. As the observing beam size becomes smaller (1.0, 0.3, 0.1$''$) for the fixed size of AME emitting region (100 pc) which, in other words, indicates that the AME emitting region becomes larger relative to the observing beams, the AME fraction observed in Band 5b (10--13GHz) increases. For the lowest redshift ($z=0.01$), the increase is significant: from $\lesssim1$\% (right panel) to a few $\times$ 10\% (left panel). The AME fraction is still negligible if $z>0.1$: the largest AME contribution from 100 pc size AME region relative to the 0.1$''$ observing beam at $z=0.1$ (i.e., 180 pc physical size) is $\lesssim 1$\% (left panel). 

\section{Discussion}
To understand the impact of AME to the radio SED for measuring SFR using thermal free-free emission, the important parameters are the density and size of the AME emitting region, relative to those of the region emitting thermal free-free emission. Simple analytical investigation in this chapter provides useful insight on our perspective of AME on the radio continuum observation using next generation radio telescopes with superb sensitivity and angular resolution.

\subsection{Impact of AME for radio continuum emission using SKA-Mid}
The highest observing frequency of SKA-Mid in Band 5b is lower than the typical peak frequency of AME ($\approx 30$ GHz). Although the overall impact of AME on the radio continuum emission observed by SKA is not very significant ($\approx 10$\% at low redshift, $z<0.1$), when observing nearby galaxies with high angular resolution ($\theta\lesssim0.1''$), multi-band radio continuum observation may still be required to characterize the shape of the radio SED and disentangle thermal free-free emission and AME in order to avoid a bias in the derived SFR one would get from a single band observation. With high angular resolution ($<0.1''$) and extreme sensitivity ($\approx 0.4$$\mu$Jy/beam 3$\sigma$ RMS for 40 hours of integration), SKA-Mid may detect AME from a high column density region ($N^{\mbox{\tiny AME}}_{\mbox{\tiny H}}=10^{23.8}$cm$^{-2}$) of interstellar medium in a galaxy at $z=1$. Therefore a high-angular resolution deep radio continuum survey to study galaxy star formation using the SKA needs an understanding of the impact of AME on the assumption of free-free dominance in the observed radio emission.     

\subsection{Implications for high-frequency ngVLA and low-frequency ALMA}
The peak frequency of AME falls within the observing windows of the high-frequency ngVLA and low-frequency ALMA. Therefore the contribution of AME to radio continuum emission observed by these two telescopes can be of the same order as the contribution from thermal free-free emission (see the right panel in Figure~\ref{fig:example}). Although it is true that AME flux density (Equation~\ref{eq:ame_flux1}) for distant galaxies is still much fainter than the sensitivity of the ngVLA and ALMA, the AME fraction in the observed radio continuum SED for nearby galaxies can be significant (more significant than in the case of SKA observations). Since the detailed ISM conditions in extragalactic star-forming regions can vary from galaxy to galaxy, one needs to be careful when interpreting the measured flux density from single-frequency radio observations and converting this to a SFR.  

\section{Summary}
In this chapter, we investigate the impact of AME on the radio continuum emission observed by the SKAO, using an approach based on an AME emissivity model combined with a prediction for the free-free emission which in turn is set by the gas surface density following \cite{kennicutt_1998}. We investigate the flux density and the fraction of AME as a function of redshift, for different frequencies within SKA-Mid Band 5b, resolutions, and column densities. Our conclusions are as follows.

\begin{enumerate}
\item[$\bullet$] The significance of the AME is determined by the size of AME region relative to the observing beam and the ISM hydrogen column density.
\item[$\bullet$] AME does not affect the radio SED significantly for distant galaxies unless there is a high density local ISM observed at extremely sensitive and high angular resolution.
\item[$\bullet$] High-angular resolution observation of nearby galaxies resolving individual star-forming regions may need multi-frequency observations to avoid a potential bias in the measure of star formation rate that is based on a single frequency.
\item[$\bullet$] Although it is challenging, detecting AME in distant galaxies by taking advantage of the AME peak frequency shift to the SKA-Mid Band 5b would be potentially very interesting as it will shed light on the evolution of small grains.
\end{enumerate}

\bibliographystyle{abbrvnat-maxbibnames4}
\bibliography{chapter} % if your bibtex file is called example.bib

\end{document}